%% file: main.tex
\documentclass[pra,twocolumn,eqsecnum,nofootinbib]{revtex4}
\include{jm-tex-macros-public-god-paper}
\usepackage{lipsum}

\begin{document}
%\maketitle
%\section{}
%\subsection{}

%\title{Brief Article}
%\author{The Author}
%\date{}							% Activate to display a given date or no date

%\title{Evidence for the existence of God from Nishimori's cat}
%\title{Decoherence, Naturalness and the existence of God}
%\title{Apparent violations of Naturalness in decoherence-induced transitions as evidence for the existence of God}
\title{Nishimori's self-tuning as evidence for the existence of God}

\author{John McGreevy}
\author{Tarun Grover}

\affiliation{Department of Physics\\University of California San Diego}

\begin{abstract}

%The smallness of the cosmological constant and of the mass of the Higgs compared to reasonable energy scales 
Apparent violations of Naturalness may be explained by positing the existence of an omniscient but disinterested and possibly fallible Observer who regularly performs von Neumann measurements on us (and everything else).  We comment briefly on the implications for the construction of scalable quantum computers.

\end{abstract}

%\today
\date{April 1, 2024}

\maketitle

%\tableofcontents

%\section{Introduction}

Field theories arise in many contexts.  
A somewhat new situation governed by field theory \cite{Dennis:2001nw,Wang:2002ph,lee2022measurement,Zhu:2022bpk,lee2023quantum,Fan:2023rvp,Bao:2023zry} is the critical behavior of phase transitions induced by decoherence.
In this context, a surprise \cite{Dennis:2001nw,Wang:2002ph,Zhu:2022bpk,lee2022measurement,Fan:2023rvp,lee2023quantum,chen2023separability,su2024tapestry,lyons2024understanding} is that apparently multicritical field theories can arise by tuning a single parameter.  
From the point of view of an effective field theorist studying this system from within, this would appear to be a gross violation of Wilson-'t Hooft naturalness \cite{tHooft:1979rat}.
%(for some perspectives on this concept, see \eg~ \cite{polchinski2006cosmological, Koren:2020pio, Craig:2022eqo}).

In this Letter, dated April 1, 2024, we propose that a similar mechanism could explain the smallness of prominently relevant operators in the Standard Model.  The only catch is that it requires the existence of an omniscient observer watching over everything, in the sense of making regular local measurements of our degrees of freedom.  We are compelled to adopt a name for such an Observer that is already used in the literature (see, for example, \cite{bible,hamilton1969mythology})\footnote{For example, Proverbs 15:3, ``In every place are the eyes of Yahweh, observing the bad and the good.'' [Trans: D.~Arovas].
%``... God knows all that is in the heavens and on the earth; God is Knowing of all things,"
%Quran 49:16
%[I also want to cite something from the Vedas about omniscience.] 
We note, however, that our mechanism does not give any evidence for a {\it unique god}.  Some sort of loosely-affiliated team of Observers \cite{hamilton1969mythology} could produce the same phenomenology.}.  However, we find that in contrast with the existing literature, consistency with the data suggests this Observer to be fallible.

The Standard Model has two prominent relevant operators that carry no symmetry quantum numbers: the quadratic invariant of the Higgs field, whose coefficient determines its mass, and the identity operator, whose coefficient determines the value of the cosmological constant.  Much ink has been spilled discussing the fact that the couplings to these operators are observed to be quite small compared to microscopic energy scales 
(see for example \cite{weinberg-cc, polchinski2006cosmological, Bousso:2007gp, Burgess:2013ara, Penco:2020kvy, Koren:2020pio, Hebecker:2020aqr, Craig:2022eqo} and the many references therein).

A common point of view in the community of people who study lattice field theory is that continuum field theories are associated with continuous phase transitions.  Tuning to the phase transition is required to approach a continuum limit, where the correlation length is much longer than the lattice spacing.  Such a procedure works in practice when the critical field theory has a single relevant operator.  Multicritical theories, with multiple relevant operators, are much harder to realize this way (unless the relevant operators break some symmetry)\footnote{We note here that some models do exhibit critical {\it phases}, 
such as the 2d XY model 
and 3+1d abelian gauge theory.  This would work just as well for providing a continuum limit. \label{footnote-one}}.

Thus, from the point of view of someone who believes that field theories arise as a long-wavelength approximation of lattice models (a point of view common in the condensed matter community (see \eg~\cite{Wen_book})), a further interesting question one might hope to address is the following.  
Given that continuum quantum field theory seems to provide an excellent description of particle physics at many length scales, what is the mechanism by which the correlation length is made large, compared to some microscopic  scale, such as the Planck length?  

Here we will not be quite so ambitious as to answer this question directly.  
However, we do suggest a mechanism by which it can be tied to the aforementioned Naturalness problems.  
We propose that the massless Standard Model could be realized as the critical theory for a decoherence-induced phase transition analogous to those mentioned above.  
The basic idea is that although the field theories describing such transitions may have two relevant operators, only one parameter needs to be tuned to reach the critical point.  
There are two scenarios in which this might be useful.
In scenario $F$, we ignore symmetries, so that chiral symmetry cannot be used to explain the masslessness of the Standard Model fermions.  Then the two relevant parameters could be\footnote{As we explain below, the masses of the gauge bosons could also be included with the fermion masses.} $m_f$, the masses of the quarks and leptons, and $m_H^2$, the Higgs mass.  We suppose that for some reason, which we do not explain, some almighty power has chosen to tune one parameter $m_f \to 0$, in order to generate a field theory with a large correlation length; the tuning of the second parameter comes for free, as we explain below.

Alternatively, in scenario $\Lambda$, we suppose that for some other reason (such as chiral symmetry), our whole phase diagram involves the massless fields of the Standard Model.  Then the idea is that the mechanism we describe could tie together the couplings of the two relevant operators, $\Lambda$ and $m_H^2$.  
We leave it as an exercise for the reader to determine which scenario is more speculative.

\newpage
{\bf Decoherence as a consequence of measurement.}
The effect of decoherence on a density matrix can be captured by the following map: 
\be\rho \to \sum_\alpha K_{\alpha} \rho K^{\dagger}_{\alpha},\ee where $\{K_\alpha\}$ are a set of `Kraus operators' that satisfy $\sum_{\alpha }K^{\dagger}_{\alpha} K_\alpha = 1$ so that the map preserves the trace of the density matrix. We now briefly review how decoherence can be realized by locally coupling to ancillas and then tracing out the ancillas. In the story we want to tell, we are forced to trace out the ancillas because we do not know the Mind of God.  

Consider, for purposes of exposition, a single qubit.  To realize the (phase-damping) channel $ \rho \to p \rho + (1-p) Z \rho Z $, we couple the qubit to an ancilla qubit.  Starting in a product state $ \rho \otimes \ket{0}\bra{0}$, evolve the combined system by the Hamiltonian $H = J Z Z_a $ for a time $t$. The result is 
\be 
e^{ - \ii t J Z Z_a } \rho \otimes \ketbra{0}{0} e^{ \ii tJ  Z Z^a} 
%= \( \cos(tJ) \Ione +  \sin(tJ) Z Z^a  \) \rho \otimes \ket{0} \bra{0}  
%\( \cos (tJ) \Ione - \sin(tJ) Z Z^a  \)
. \ee
Using the identity $e^{ - \ii t J Z Z_a }  =  \cos(tJ) \Ione +  \sin(tJ) Z Z^a  $ and 
tracing out the ancilla then gives 
\be  \rho \to \cos^2 (tJ) \rho + \sin^2(tJ) Z \rho Z ~;\ee
this is of the indicated form with $p = \cos^2(tJ)$.  
So the probability of acting with the nontrivial Kraus operator $Z$ is determined by the strength of the coupling $J$.  

The analogous many-body channel is realized by coupling each qubit to its own ancilla, via $ H = \sum_i J_i Z_i Z_i^a $.  
These describe the Eyes of God.  
By varying the couplings $J_i$, we can vary the probabilities with which we act with a given $Z_j$.
We do not comment on whether God remembers the results of the measurements They make, or on the question of whether They can run out of fresh ancillas\footnote{The limited availability of fresh ancillas could provide a limit on the possible number of efoldings of inflation.}.  For our purposes, what matters is that \textit{we} don't know the outcome of these measurements. 

Based on common opinion, it may seem natural to assume that God does not make any measurement errors\footnote{It is surprisingly difficult to find a reference in the literature for this statement.}. However, we will see that
allowing the omniscient Observer in our scenario to be fallible naturally leads to the existence of a time direction in the underlying description.

%However, we will see that in fact the omniscient observer in our scenario {\it must be fallible} in order to allow for coherence in the time direction.   In the absence of measurement errors, the critical field theory is only $3+0$ dimensional \cite{Dennis:2001nw}.  

%[Comment on the choice of basis in which the measurements are done, \ie~phase damping versus bit flip channel.]

\begin{figure}
$$ \includegraphics[width=.3\textwidth]{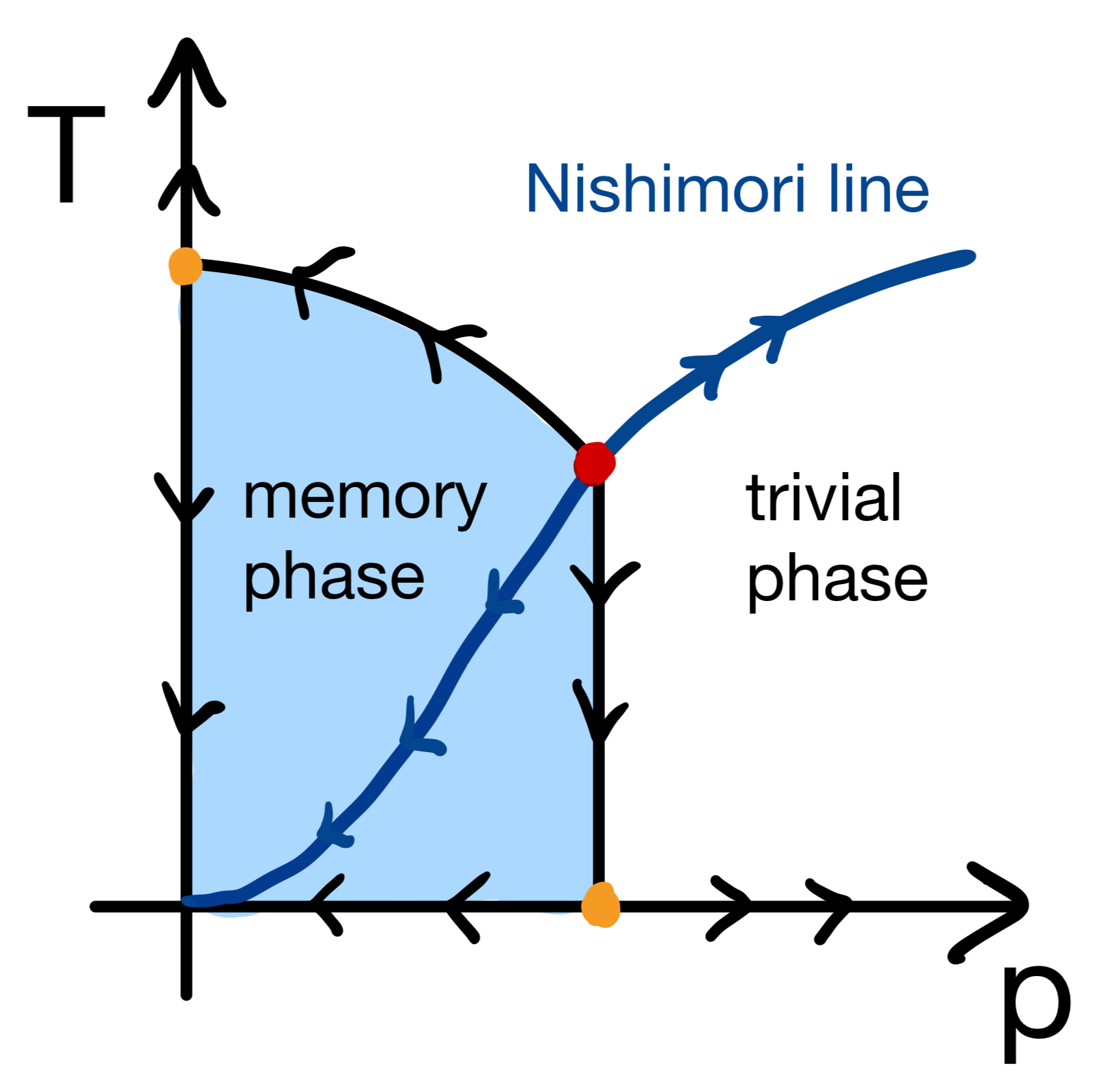}$$
$$ \includegraphics[width=.2\textwidth]{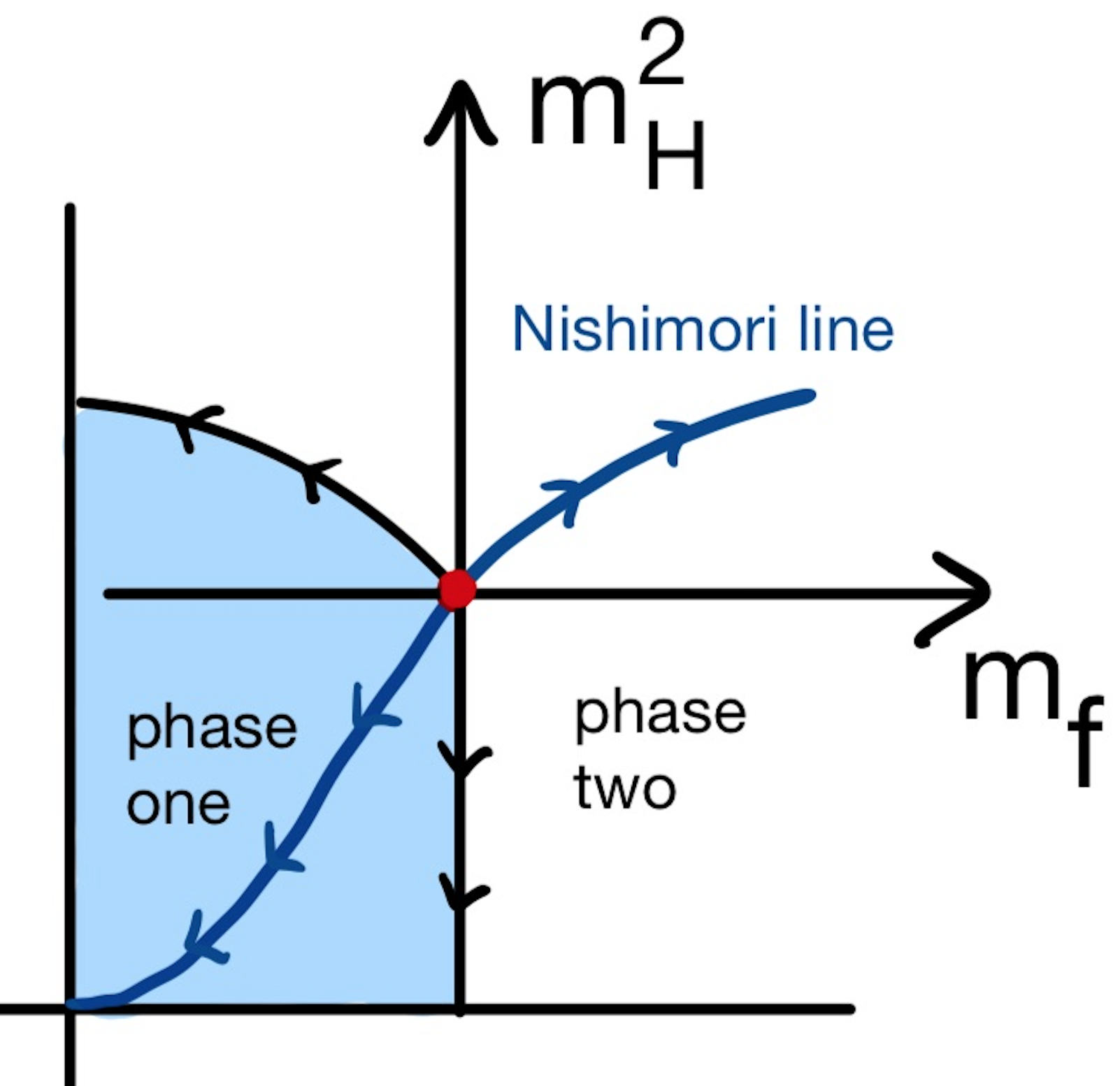}~~\includegraphics[width=.2\textwidth]{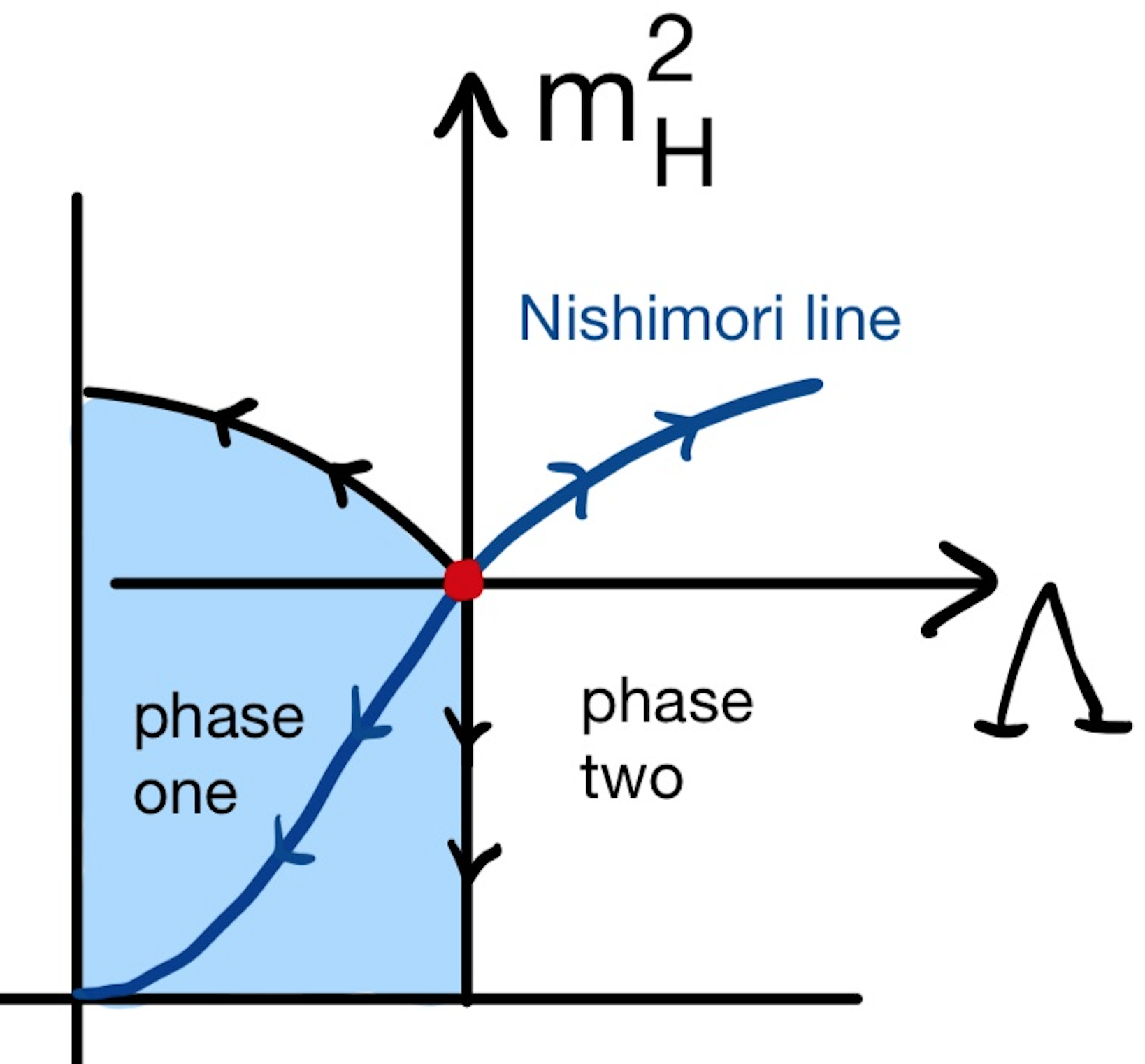}$$
\caption{\label{fig:nishimori-line}
The top figure depicts the phase diagram of a model such as the random-bond Ising model, as a function of temperature and disorder strength.  The arrows indicate the direction of RG flow, towards the infrared.  
The orange circles thus represent the generic critical behavior.
When this model arises as a result of forgetful measurements, the system lies on the Nishimori line, determining $T$ in terms of $p$.  Thus, even though the critical point (red circle) has two relevant operators, only one parameter must be tuned to reach it.  
The other two pictures depict scenarios $F$ and $\Lambda$ described in the main text for how this mechanism might provide some theological insight.
}
\end{figure}

{\bf Reminder about Nishimori's self-tuning phenomenon.}
Here we review the phenomenon of self-tuning in critical field theories governing decoherence-induced phase transitions.

For our purposes, the key fact  \cite{Dennis:2001nw, Wang:2002ph, lee2022measurement, Zhu:2022bpk,Chen:2023tfg} about decoherence-induced transitions (at least in known examples) is that one is put directly on the Nishimori line 
of a statistical mechanics model 
\cite{nishimori1981internal,nishimori1993optimum}, where the probability of a given disorder realization $C$ is proportional to the partition function of the system with that disorder realization $Z(C)$.
This happens because the probability here is determined by the Born rule. As an example, consider the 2d toric code under bit-flip or phase-flip decoherence. Under either of these channels, when the measurements are flawless, the statistical mechanics model corresponds to the random bond Ising model (RBIM) along the `Nishimori line'. Recall that in the phase diagram of the RBIM, there are generically two relevant parameters: the temperature $T$, and the relative probability, $p/(1-p)$, for a bond to be $+J$ or $-J$. Along the Nishmori line, these two parameters are related via $\tanh(1/T) = 1-2p$, and one undergoes a phase transition from a ferromagnetic phase (low $T$, low $p$ regime) to a paramagnetic phase (high $T$, high $p$ regime). Generically, both $T$ and $p$ are relevant parameters at this Nishimori critical point. However, when this critical point is realized in the context of decoherence-induced transitions, there is a single tuning parameter, namely, the rate $p$ of decoherence, while the temperature $T$ is determined by the aforementioned relation $\tanh(1/T) = 1-2p$.
From the point of view of effective field theory, this is rather surprising. One is tuning a single parameter to reach a multicritical point which has two relevant directions (see Fig.~\ref{fig:nishimori-line}). The phase diagram does have two other fixed-points that have a single relevant direction, but they are not accessible in this setup due to the Born rule.
We expect there to be many other examples of this phenomenon, realizing the Nishimori line of other statistical models.
%[MORE EXAMPLES]

We now encounter an important subtlety.  If the measurements are performed flawlessly on the same operator at all times, the critical field theory that emerges at such a measurement-induced transition is a Euclidean theory in $d$ spatial dimensions, with no time coordinate.  This would not suffice for our purposes unless the Observer lived in an extra time dimension, and our time dimension arose via some process of Wick rotation among the $d$ dimensions of the field theory.  Though there are suggestions along these lines in the literature 
(see \eg~\cite{god-and-time}),
%[cite bible passages about how "God lives outside of time"], 
we find the following observation more appealing.  
Consider the case where the Observer makes `measurement errors' at a rate $q$.  
This means that, although the Observer is trying to measure an operator $\CO$, with probability $q$, They will instead measure an operator $\CO'$.  
The description of such a system \cite{Dennis:2001nw,Wang:2002ph} now involves a time dimension.  
For example, when performing imperfect measurements on a 2d toric code, one obtains a 3d random-plaquette gauge model.
If one allows for imperfect measurements, one is generically not on the Nishimori line. 
However, when $p=q$ two things happen simultaneously: first, the system has a symmetry that rotates time and space dimensions, a necessary step for Lorentz invariance.  Secondly, we are placed on the Nishimori line of the phase diagram of a $d+1$ dimensional system\footnote{We cannot exclude the unnatural-seeming possibility that the Observer is intentionally choosing a measurement schedule that appears to us to result from measurement errors, and thus cannot definitively conclude that They must be fallible.}.

We must discuss a few additional subtleties about our proposal.  
A local finite depth channel ${\cal E}$, such as that describing local forgetful measurements, cannot produce long-range correlations from a short-range correlated state, 
since, for such a channel, $ \tr {\cal E}(\rho) \CO_1 \CO_2  = \tr \rho { \cal E}^\dagger(\CO_1\CO_2)$ is a correlation function of local operators. 
%Thus, it seems that scenario $F$ cannot be realized if we restrict to an Observer who makes only {\it local} measurements.   
%This appears to be consistent with various statements in the literature [cite bible passages about God being everywhere...].  
%$ \tr {\cal E}(\rho) \CO_1 \CO_2  = \tr \rho { \cal E}^\dagger(\CO_1\CO_2) 
%= \tr \rho { \cal E}^\dagger(\CO_1){ \cal E}^\dagger(\CO_2)$, 
Because of this, we know that the density matrix of the system which is subjected to the measurements is not directly described by the critical theory  \cite{Fan:2023rvp}.  Indeed, starting with the (gapped) toric code groundstate and doing local measurements for a finite time cannot produce long-range correlations (that is, \tr $\CE(\rho) \CO_1 \CO_2$ is not a power-law in the separation between the two operators).
What exactly is the relation between these two systems?  
One way to understand the relation is by considering observables that are \textit{non-linear} in the density matrix, e.g., the von Neumann entropy $=-\tr(\rho \log(\rho))$ or mixed-state entanglement measures such as entanglement negativity \cite{lee2023quantum,Fan:2023rvp}. The singular characteristics of such non-linear observables are directly related to the underlying critical theory. Yet another way is to seek an `optimal decomposition' of the density matrix, as described in \cite{chen2023separability}.  There it is shown that the density matrix of the decohered system can be written as a mixture of pure states
\be\label{eq:optimal-decomposition} \rho = \sum_i p_i \ketbra{\psi_i}{\psi_i}\ee
{\it each of which} undergoes the decoherence-induced phase transition.  That is, for each value of the tuning parameter $p$, 
each $\ket{\psi_i}$ 
is a representative of the phase at that value of $p$, 
with short-ranged correlations away from the transition, and a diverging correlation length at $p_c$.  
An alternative idea to realize scenario $F$, then, is to think of the ensemble \eqref{eq:optimal-decomposition} as an incoherent analog of the many-worlds picture of quantum mechanics -- we are realized as one element of this ensemble.  
%From this point of view, the 

{\bf The critical Standard Model.}  
Let us explain scenario $F$ in a bit more detail.  The idea has some ingredients in common with the phenomenon of a deconfined quantum critical point (DQCP) \cite{Senthil:2004fuw, Senthil:2003eed}.    
This is a phase transition separating two ordinary (usually gapped) phases which is described by a gauge theory.  On both sides of the transition, the gauge degrees of freedom are confined or Higgsed, and only at the transition are there massless gauge fields (and usually also critical matter degrees of freedom).  
So there do exist transitions where only at the transition are the matter fields massless, while the gauge fields are also absent away from the critical point.  
A remarkable thing about a DQCP is that the whole gauge theory only emerges at the critical point, but with a single tuning\footnote{Though see recent literature for discussion of the possibility that the canonical example of this phenomenon may have a second relevant operator \cite{Zhao:2020ydp,Lu:2021ucu,DEmidio:2024ggo}.  Perhaps experiments trying to realize this theory should also use the Nishimori mechanism.}.
The $m_f$ in Fig.~\ref{fig:nishimori-line} is thus a proxy for the tuning parameter across such a transition.  

In 3+1 dimensions,  it is believed to be difficult to realize the phenomenon of deconfined quantum criticality \cite{hosur2010chiral}.  
The natural extension of the canonical Neel-VBS transition to three dimensions instead is described by a an intermediate critical {\it phase}.  
This can happen because (compact) $\gU(1)$ gauge theory in 3+1 dimensions has a deconfined phase, even in the absence of matter (unlike in 2+1d \cite{Polyakov:1975rs, Polyakov:1976fu}).  
We have two points to make about this.  Firstly, it may be that this apparent difficulty is just a property of the simplest examples.  
More importantly, as mentioned in footnote \ref{footnote-one}, a critical phase would work just as well for explaining the fact that we live in the continuum.

The suggestion that the massless Standard Model could govern a continuous phase transition is not too radical.  Critical points between unspectacular phases governed by non-abelian gauge theories are known \cite{Gazit:2018vsa, Bi:2018xvr,Bi:2019ers, Ma_2020, Zou:2020hqi}.  But what about critical points governed by {\it chiral} gauge theories?  
%[Are there any proposed examples in the literature?]
What are the two phases separated by the putative Standard Model DQCP? See \cite{Wang:2021hob, Wang:2021vki, You:2022egv} for other connections between the Standard Model and the DQCP phenomenon.
We leave these problems as exercises for the reader.

{\bf Possible observable consequences.}  
The critical field theories realized by decoherence-induced phase transitions differ in several ways from field theories with more familiar origins. 
First, translation invariance only holds on average.  The specific time series of measurements explicitly breaks translation symmetry, only the ensemble is translation invariant.   As a result, an observer such as ourselves living in such a system should see violations of momentum conservation, on time-scales shorter than the time it takes
God's measurements to sweep across the system.  Although the total measurement rate is fixed to be near the critical rate, presumably more frequent but weaker measurements could allow this sweeping time to be short enough to have evaded detection thus far.  

Secondly, such a quantum field theory is not an isolated system.  Quantum mechanical probability can leak out of the system, and not all of the familiar consequences of unitarity need hold.  It is possible that such constraints already rule out both scenarios.  One could imagine seeing violations of the optical theorem at high energies.  One could imagine seeing failures of quantum interference experiments because one of the participants is measured by the Observer during the course of the experiment.  Assuming the required measurement rate and strength are small enough that the resulting violations do not already falsify our scheme, we can expect that as we do experiments involving quantum coherence over larger and larger systems, eventually we will encounter such events.  The Observer's measurement rate represents an 
intrinsic source of decoherence against which the construction of any quantum computer must fight.  
Recall the remarkable result that there exist error-correcting codes 
with the property that they allow for the protection of a quantum computation from errors below a certain threshold rate, $p_\star$ (per qubit, per clock cycle), which depends on the code \cite{shor1996fault,aharonov1997fault,kitaev2003fault,knill1998resilient,preskill1998reliable,terhal2015quantum}.  If the Observer's measurement rate (when translated into a rate per qubit per clock cycle) is larger than $p_\star$, then the code in question cannot be useful.  

In experiments, how can we distinguish between decoherence due to ordinary couplings to a terrestrial environment, and decoherence due to being watched by God?  It seems that the only solution is to attempt to isolate the apparatus in the usual ways, by making a good vacuum, by making it cold, by burying it underground, while maximizing its volume.  We can envision a future where experiments which are now used to detect neutrinos and to try to detect dark matter will be repurposed for direct detection experiments looking for the Eyes of God.  

{\bf Discussion.}  
Above we have suggested a possible mechanism for naturally realizing a version of the Standard Model where relevant operators are fine-tuned to zero.
However, the cosmological constant and the Higgs mass are not zero.  From the viewpoint of our theory (in scenario $\Lambda$), the Electroweak Scale and the nonzero cosmological constant value are corrections due to being slightly off-criticality.  That is, they arise because God is not measuring at exactly the critical rate.
A second issue is that a critical version of the Standard Model would also involve tuning the QCD gauge coupling to zero -- it is a marginally relevant perturbation of the free fixed point. 
The electroweak and hypercharge gauge couplings are also marginally (ir)relevant operators.  The same is true of the Higgs self-coupling and the Yukawa couplings.  One can imagine that there is a critical point somewhere in this space of couplings, at least in the $4-\epsilon$ expansion.
Presumably the deviation of the couplings from this fixed point could be explained similarly.
We do not comment on neutrino masses.

Existing studies of decoherence-induced transitions have been performed on lattice models.  
To apply these ideas to scenario $F$, where we hope to couple the explanation of  the continuum limit with the smallness of the Higgs mass,
we must assume a lattice regularization of the Standard Model.  
Lattice regularizations of many Lorentz-invariant gauge theories are by now well-understood.  However, chiral gauge theories such as the SM pose obstacles for such a regularization \cite{Nielsen:1981hk} that have not yet been surmounted.  Currently the best idea is to begin with a doubled version of the theory and rely on a mechanism of Symmetric Mass Generation to give a large mass to the mirror degrees of freedom \cite{Eichten:1985ft,Zeng:2022grc,Wang:2022ucy}.  We assume that such a process can be made to work.  
Scenario $\Lambda$ may take place entirely in the continuum; this would require a continuum version of the decoherence channel.

It is not yet clear {\it which} or {\it how many} relevant operators will be forbidden by the Born rule. 
%The Born rule imposes a restriction on the couplings...
It would be extremely interesting to find examples where out of $m$ relevant operators only $n < m$ lie on the Nishimori line, for other values of $n$ and $m$.  

We hasten to point out that in the known examples of the Nishimori tuning phenomenon, the couplings that are automatically tuned away are perturbations involving spatial disorder.  
%The critical theories describing decoherence-induced transitions may not even enjoy lattice translation symmetry.  
%[Discuss the translation symmetry on-average.]

We note that our suggestion does not involve gravity, which is a difficulty for scenario $\Lambda$.  Whether the coefficient of the identity operator in the action of a QFT without gravity is meaningful is open to debate.

We also have not explained how the Observer chose Their measurement rate, or why They chose it to be close to the critical value.  

The apparent presence of fine-tuning in the constants of Nature has previously been interpreted as evidence for the existence a higher power (see \eg~\cite{sawyer2009calculating, deutsch2011beginning} for a discussion).  However, such an interpretation requires an imputation of particular goals on the part of that higher power, such as some desire to create complex organisms like ourselves, which can only be described as hubristic.  The mechanism described in this paper, while a bit more subtle, requires no such goals on the part of the higher power, only Their attention.

%[I put this dummy text here because the footnote won't appear if the text is only one page long.] \lipsum[2-4]

%============
% end of the main text
%=============
%\vfill\eject
{\bf Acknowledgment.} 
We are grateful to 
Dan Arovas for sharing some of his biblical expertise.
JM thanks Shamit Kachru for bringing \cite{sawyer2009calculating} to his attention.
The funding agencies that support our other work are not responsible for funding the writing of this paper.  We did it in our spare time, we promise.  

%This work was supported in part by
%funds provided by the U.S. Department of Energy
%(D.O.E.) under cooperative research agreement 
%DE-SC0009919, 
%%by the University of California Laboratory Fees Research Program, grant LFR-20-653926,
%and by the Simons Collaboration on Ultra-Quantum Matter, which is a grant from the Simons Foundation (652264, JM).
%JM received travel reimbursement from the Simons Foundation;
%the terms of this arrangement have been reviewed and approved by the University of California, San Diego in accordance with its conflict of interest policies. 

%\appendix
%\renewcommand{\theequation}{\Alph{section}.\arabic{equation}}

\phantomsection\addcontentsline{toc}{section}{References}
\bibliographystyle{ucsd}
\bibliography{refs} 
%\phantomsection\addcontentsline{toc}{section}{References}
%%\bibliography{collection} 
% \bibliography{\jobname}
%% \bibliographystyle{alpha}
%\bibliographystyle{ucsd}
% \openout0= \jobname.bib
% \write0{
%   %%%%%%%%%%%%%%%%%%%%%%%%%%%%%%%%
%   %%% more bib entries go here %%%
%   %%%%%%%%%%%%%%%%%%%%%%%%%%%%%%%%
% }

\end{document}

%% file: jm-tex-macros-public-god-paper.tex
\usepackage{amssymb}
\usepackage{amsmath,bm}
\usepackage{amssymb}
\usepackage{graphicx}
\usepackage{amsfonts}         
\usepackage{fancybox}

\usepackage[usenames,dvipsnames]{xcolor}%http://en.wikibooks.org/wiki/LaTeX/Colors

\definecolor{Mathematica1}{rgb}{0.368417, 0.506779, 0.709798}
\definecolor{Mathematica2}{rgb}{0.880722, 0.611041, 0.142051}
\definecolor{Mathematica3}{rgb}{0.560181, 0.691569, 0.194885}

\usepackage[usenames,dvipsnames]{xcolor}%http://en.wikibooks.org/wiki/LaTeX/Colors

\usepackage[pagebackref,  % this puts links to the page numbers where refs appear
bookmarks={false}, pdfauthor={John McGreevy, Tarun Grover}, pdftitle={Please note the date of the paper!}]{hyperref}
\hypersetup{colorlinks=true, 
linkcolor=Mathematica1,
citecolor=Mathematica3,
filecolor=OliveGreen, 
urlcolor=Mathematica2,
filebordercolor={.8 .8 1}, 
urlbordercolor={.8 .8 0}}%http://en.wikibooks.org/wiki/LaTeX/Hyperlinks
\usepackage{soul}%strike through text \st{} .  not in math mode.
\setstcolor{Red}
\definecolor{darkgreen}{rgb}{0,0.4,0}
\definecolor{darkred}{rgb}{0.4,0,0}
\definecolor{darkblue}{rgb}{0,0,0.4}
\definecolor{lightblue}{rgb}{.6,.6,0.9}

\definecolor{uglybrown}{rgb}{0.8,  0.7,  0.5}

\definecolor{palatinatepurple}{rgb}{0.41, 0.16, 0.38}
\definecolor{celebrationcolor}{rgb}{0.75,  0.0,  0.9}

\definecolor{Mathematica1}{rgb}{0.368417, 0.506779, 0.709798}
\definecolor{Mathematica2}{rgb}{0.880722, 0.611041, 0.142051}
\definecolor{Mathematica3}{rgb}{0.560181, 0.691569, 0.194885}

\definecolor{shadecolor}{rgb}{0.90,0.90,0.90}
\definecolor{DVcolor}{rgb}{0.95,  0.5,  0.2}
\definecolor{lightbluemuons}{rgb}{0.0,.65,1.0}

\definecolor{chartreuse}{rgb}{0.70, 1.00, 0.00}

\usepackage{tikz}
\usetikzlibrary{shapes}
\usetikzlibrary{trees}
\usetikzlibrary{snakes}
\usetikzlibrary{matrix,arrows} 				% For commutative diagram
\usetikzlibrary{positioning}				% For "above of=" commands
\usetikzlibrary{calc,through}				% For coordinates
\usetikzlibrary{decorations.pathreplacing}  % For curly braces
\usepackage[tikz]{bclogo} 					% For cute logo boxes
\usepackage{pgffor}							% For repeating patterns

\usetikzlibrary{decorations.markings}
\usetikzlibrary{intersections}				% For interesections

\usetikzlibrary{arrows,decorations.pathmorphing,backgrounds,positioning,fit,petri,automata,shadows,calendar,mindmap, graphs}
\usetikzlibrary{arrows.meta,bending}

\tikzset{
    vector/.style={decorate, decoration={snake}, draw},
    fermion/.style={postaction={decorate},
        decoration={markings,mark=at position .55 with {\arrow{>}}}},
    fermionbar/.style={draw, postaction={decorate},
        decoration={markings,mark=at position .55 with {\arrow{<}}}},
    fermionnoarrow/.style={},
    gluon/.style={decorate,
        decoration={coil,amplitude=4pt, segment length=5pt}},
    scalar/.style={dashed, postaction={decorate},
        decoration={markings,mark=at position .55 with {\arrow{>}}}},
    scalarbar/.style={dashed, postaction={decorate},
        decoration={markings,mark=at position .55 with {\arrow{<}}}},
    scalarnoarrow/.style={dashed,draw},
	vectorscalar/.style={loosely dotted,draw=black, postaction={decorate}},
}

\def\centerarc[#1](#2)(#3:#4:#5)% Syntax: [draw options] (center) (initial angle:final angle:radius)
    { \draw[#1] ($(#2)+({#5*cos(#3)},{#5*sin(#3)})$) arc (#3:#4:#5); }

\usepackage{enumitem}

\usepackage{slashed}

\usepackage[font=small,labelfont=bf]{caption}
\DeclareCaptionFont{tiny}{\tiny}
\usepackage{epstopdf}
\usepackage{wasysym}

\usepackage[yyyymmdd,hhmmss]{datetime}   
\usepackage{framed}  % needed for \begin{shaded}
 \usepackage{mdframed}
 \usepackage{wrapfig}
 \usepackage{yfonts}  

\def\ketbra#1#2{ | #1 \rangle\hskip-2pt\langle #2|}

\newmdenv[%
        backgroundcolor=lightgray,
    linecolor=black,
    outerlinewidth=2pt,
]{boxedandshaded}

\numberwithin{equation}{section}

\newlength{\extraspace}
\newlength{\extraspaces}
\def\be{\begin{equation}}
\def\ee{\end{equation}}

\newcommand{\bea}{\begin{eqnarray}}
\newcommand{\eea}{\end{eqnarray}}

\def\tr{{\rm tr}}

\def\bra#1{\left\langle#1\right|}
\def\ket#1{\left|#1\right\rangle}

\def\CE{{\cal E}}

\def\CO{{\cal O}}%AEL

\def\II{\relax{I\kern-.10em I}}

\def\IB{\relax{\rm I\kern-.18em B}}

\def\ID{\relax{\rm I\kern-.18em D}}
\def\IE{\relax{\rm I\kern-.18em E}}
\def\IF{\relax{\rm I\kern-.18em F}}
\def\IG{\relax\hbox{$\inbar\kern-.3em{\rm G}$}}
\def\IGa{\relax\hbox{${\rm I}\kern-.18em\Gamma$}}
\def\IH{\relax{\rm I\kern-.18em H}}
\def\II{\relax{\rm I\kern-.18em I}}
\def\IK{\relax{\rm I\kern-.18em K}}
\def\inbar{\,\vrule height1.5ex width.4pt depth0pt}

\def\simgt{\hskip0.05in\relax{ 
\raise3.0pt\hbox{ $>$
{\lower5.0pt\hbox{\kern-1.05em $\sim$}} }} \hskip0.05in}

\def\lp10{\ell_p^{10}}
\def\lp11{\ell_p^{11}}
\def\R11{R_{11}}

\def\frac#1#2{{#1 \over #2}}

\def\Ione{\hbox{$1\hskip -1.2pt\vrule depth 0pt height 1.53ex width 0.7pt
                  \vrule depth 0pt height 0.3pt width 0.12em$}}

\newdimen\tableauside\tableauside=1.0ex
\newdimen\tableaurule\tableaurule=0.4pt
\newdimen\tableaustep
\def\phantomhrule#1{\hbox{\vbox to0pt{\hrule height\tableaurule width#1\vss}}}
\def\phantomvrule#1{\vbox{\hbox to0pt{\vrule width\tableaurule height#1\hss}}}
\def\sqr{\vbox{%
  \phantomhrule\tableaustep
  \hbox{\phantomvrule\tableaustep\kern\tableaustep\phantomvrule\tableaustep}%
  \hbox{\vbox{\phantomhrule\tableauside}\kern-\tableaurule}}}
\def\squares#1{\hbox{\count0=#1\noindent\loop\sqr
  \advance\count0 by-1 \ifnum\count0>0\repeat}}
\def\tableau#1{\vcenter{\offinterlineskip
  \tableaustep=\tableauside\advance\tableaustep by-\tableaurule
  \kern\normallineskip\hbox
    {\kern\normallineskip\vbox
      {\gettableau#1 0 }%
     \kern\normallineskip\kern\tableaurule}%
  \kern\normallineskip\kern\tableaurule}}
\def\gettableau#1 {\ifnum#1=0\let\next=\null\else
  \squares{#1}\let\next=\gettableau\fi\next}

\def\({\left(}
\def\){\right)}

\def\ii{{\bf i}}

\def\lsim{\mathrel{\mathstrut\smash{\ooalign{\raise2.5pt\hbox{$<$}\cr\lower2.5pt\hbox{$\sim$}}}}}
\def\gsim{\mathrel{\mathstrut\smash{\ooalign{\raise2.5pt\hbox{$>$}\cr\lower2.5pt\hbox{$\sim$}}}}}

\def\overleftrightarrow#1{\vbox{\ialign{##\crcr
     $\leftrightarrow$\crcr\noalign{\kern-0pt\nointerlineskip}
     $\hfil\displaystyle{#1}\hfil$\crcr}}}
     
     \def\overleftarrow#1{\vbox{\ialign{##\crcr
     $\leftarrow$\crcr\noalign{\kern-0pt\nointerlineskip}
     $\hfil\displaystyle{#1}\hfil$\crcr}}}

\def\eg{{\it e.g.}}

\def\gU{\textsf{U}}

\newif{\ifeq}           % defines a new condition @eq tested by the conditional \ifeq
\eqtrue                 % if uncommented, declares @eq to be true
\newcounter{lecturecounter}

%% file: main.bbl
\begingroup\raggedright\begin{thebibliography}{10}

\bibitem{Dennis:2001nw}
E.~Dennis, A.~Kitaev, A.~Landahl, and J.~Preskill, ``{Topological quantum
  memory},'' {\em J. Math. Phys.} {\bf 43} (2002) 4452--4505,
  \href{http://arxiv.org/abs/quant-ph/0110143}{{\tt quant-ph/0110143}}.

\bibitem{Wang:2002ph}
C.~Wang, J.~Harrington, and J.~Preskill, ``{Confinement Higgs transition in a
  disordered gauge theory and the accuracy threshold for quantum memory},''
  {\em Annals Phys.} {\bf 303} (2003) 31--58,
  \href{http://arxiv.org/abs/quant-ph/0207088}{{\tt quant-ph/0207088}}.

\bibitem{lee2022measurement}
J.~Y. Lee, W.~Ji, Z.~Bi, and M.~Fisher, ``Measurement-Prepared Quantum
  Criticality: from Ising model to gauge theory, and beyond,'' {\em arXiv
  preprint arXiv:2208.11699} (2022).

\bibitem{Zhu:2022bpk}
G.-Y. Zhu, N.~Tantivasadakarn, A.~Vishwanath, S.~Trebst, and R.~Verresen,
  ``{Nishimori\textquoteright{}s Cat: Stable Long-Range Entanglement from
  Finite-Depth Unitaries and Weak Measurements},'' {\em Phys. Rev. Lett.} {\bf
  131} (2023), no.~20 200201, \href{http://arxiv.org/abs/2208.11136}{{\tt
  2208.11136}}.

\bibitem{lee2023quantum}
J.~Y. Lee, C.-M. Jian, and C.~Xu, ``Quantum Criticality Under Decoherence or
  Weak Measurement,'' {\em PRX Quantum} {\bf 4} (Aug, 2023) 030317,
  \href{https://link.aps.org/doi/10.1103/PRXQuantum.4.030317}{https://link.aps.org/doi/10.1103/PRXQuantum.4.030317}.

\bibitem{Fan:2023rvp}
R.~Fan, Y.~Bao, E.~Altman, and A.~Vishwanath, ``{Diagnostics of mixed-state
  topological order and breakdown of quantum memory},''
  \href{http://arxiv.org/abs/2301.05689}{{\tt 2301.05689}}.

\bibitem{Bao:2023zry}
Y.~Bao, R.~Fan, A.~Vishwanath, and E.~Altman, ``{Mixed-state topological order
  and the errorfield double formulation of decoherence-induced transitions},''
  \href{http://arxiv.org/abs/2301.05687}{{\tt 2301.05687}}.

\bibitem{chen2023separability}
Y.-H. Chen and T.~Grover, ``Separability transitions in topological states
  induced by local decoherence,'' \href{http://arxiv.org/abs/2309.11879}{{\tt
  2309.11879}}.

\bibitem{su2024tapestry}
K.~Su, Z.~Yang, and C.-M. Jian, ``Tapestry of dualities in decohered quantum
  error correction codes,'' {\em arXiv preprint arXiv:2401.17359} (2024).

\bibitem{lyons2024understanding}
A.~Lyons, ``Understanding Stabilizer Codes Under Local Decoherence Through a
  General Statistical Mechanics Mapping,'' {\em arXiv preprint
  arXiv:2403.03955} (2024).

\bibitem{tHooft:1979rat}
G.~'t~Hooft, ``{Naturalness, chiral symmetry, and spontaneous chiral symmetry
  breaking},'' {\em NATO Sci. Ser. B} {\bf 59} (1980) 135--157.

\bibitem{bible}
M.~D. Coogan, M.~Z. Brettler, C.~A. Newsom, and E.~Perkins, Pheme, eds., {\em
  The New Oxford Annotated Bible with Apocrypha: New Revised Standard Version}.
\newblock Oxford University Press, USA, 2010.

\bibitem{hamilton1969mythology}
E.~Hamilton and S.~Savage, {\em Mythology}.
\newblock Mentor book. Mentor, 1969.

\bibitem{weinberg-cc}
S.~Weinberg, ``The cosmological constant problem,'' {\em Rev. Mod. Phys.} {\bf
  61} (Jan, 1989) 1--23,
  \href{https://link.aps.org/doi/10.1103/RevModPhys.61.1}{https://link.aps.org/doi/10.1103/RevModPhys.61.1}.

\bibitem{polchinski2006cosmological}
J.~Polchinski, ``The Cosmological Constant and the String Landscape,'' 2006.

\bibitem{Bousso:2007gp}
R.~Bousso, ``{TASI Lectures on the Cosmological Constant},'' {\em Gen. Rel.
  Grav.} {\bf 40} (2008) 607--637, \href{http://arxiv.org/abs/0708.4231}{{\tt
  0708.4231}}.

\bibitem{Burgess:2013ara}
C.~P. Burgess, ``{The Cosmological Constant Problem: Why it's hard to get Dark
  Energy from Micro-physics},'' in {\em {100e Ecole d'Ete de Physique:
  Post-Planck Cosmology}}, pp.~149--197, 2015.
\newblock \href{http://arxiv.org/abs/1309.4133}{{\tt 1309.4133}}.

\bibitem{Penco:2020kvy}
R.~Penco, ``{An Introduction to Effective Field Theories},''
  \href{http://arxiv.org/abs/2006.16285}{{\tt 2006.16285}}.

\bibitem{Koren:2020pio}
S.~Koren, {\em {New Approaches to the Hierarchy Problem and their Signatures
  from Microscopic to Cosmic Scales}}.
\newblock PhD thesis, UC, Santa Barbara (main), 2020.
\newblock \href{http://arxiv.org/abs/2009.11870}{{\tt 2009.11870}}.

\bibitem{Hebecker:2020aqr}
A.~Hebecker, ``{Lectures on Naturalness, String Landscape and Multiverse},''
  \href{http://arxiv.org/abs/2008.10625}{{\tt 2008.10625}}.

\bibitem{Craig:2022eqo}
N.~Craig, ``{Naturalness: past, present, and future},'' {\em Eur. Phys. J. C}
  {\bf 83} (2023), no.~9 825, \href{http://arxiv.org/abs/2205.05708}{{\tt
  2205.05708}}.

\bibitem{Wen_book}
X.-G. Wen, {\em Quantum Field Theory of Many-body Systems: From the Origin of
  Sound to an Origin of Light and Electrons.}
\newblock Oxford University Press, Great Clarendon Street, Oxford OX2 6DP,
  2004.

\bibitem{Chen:2023tfg}
E.~H. Chen {\em et.~al.}, ``{Realizing the Nishimori transition across the
  error threshold for constant-depth quantum circuits},''
  \href{http://arxiv.org/abs/2309.02863}{{\tt 2309.02863}}.

\bibitem{nishimori1981internal}
H.~Nishimori, ``Internal energy, specific heat and correlation function of the
  bond-random Ising model,'' {\em Progress of Theoretical Physics} {\bf 66}
  (1981), no.~4 1169--1181.

\bibitem{nishimori1993optimum}
H.~Nishimori, ``Optimum decoding temperature for error-correcting codes,'' {\em
  Journal of the Physical Society of Japan} {\bf 62} (1993), no.~9 2973--2975.

\bibitem{god-and-time}
G.~E. Ganssle, ``God and Time,'' {\em Internet Encyclopedia of Philosophy}
  (2024) \href{https://iep.utm.edu/god-time/}{https://iep.utm.edu/god-time/}.

\bibitem{Senthil:2004fuw}
T.~Senthil, L.~Balents, S.~Sachdev, A.~Vishwanath, and M.~P.~A. Fisher,
  ``{Quantum criticality beyond the Landau-Ginzburg-Wilson paradigm},'' {\em
  Phys. Rev. B} {\bf 70} (2004), no.~14 144407,
  \href{http://arxiv.org/abs/cond-mat/0312617}{{\tt cond-mat/0312617}}.

\bibitem{Senthil:2003eed}
T.~Senthil, A.~Vishwanath, L.~Balents, S.~Sachdev, and M.~P.~A. Fisher,
  ``{Deconfined Quantum Critical Points},'' {\em Science} {\bf 303} (2004),
  no.~5663 1490--1494, \href{http://arxiv.org/abs/cond-mat/0311326}{{\tt
  cond-mat/0311326}}.

\bibitem{Zhao:2020ydp}
B.~Zhao, J.~Takahashi, and A.~W. Sandvik, ``{Multicritical deconfined
  quantum-criticality and Lifshitz point of a helical valence-bond phase},''
  {\em Phys. Rev. Lett.} {\bf 125} (2020), no.~25 257204,
  \href{http://arxiv.org/abs/2005.10184}{{\tt 2005.10184}}.

\bibitem{Lu:2021ucu}
D.-C. Lu, C.~Xu, and Y.-Z. You, ``{Self-duality protected multicriticality in
  deconfined quantum phase transitions},'' {\em Phys. Rev. B} {\bf 104} (2021),
  no.~20 205142, \href{http://arxiv.org/abs/2104.05147}{{\tt 2104.05147}}.

\bibitem{DEmidio:2024ggo}
J.~D'Emidio and A.~W. Sandvik, ``{Entanglement entropy and deconfined
  criticality: emergent SO(5) symmetry and proper lattice bipartition},''
  \href{http://arxiv.org/abs/2401.14396}{{\tt 2401.14396}}.

\bibitem{hosur2010chiral}
P.~Hosur, S.~Ryu, and A.~Vishwanath, ``Chiral topological insulators,
  superconductors, and other competing orders in three dimensions,'' {\em
  Physical Review B} {\bf 81} (2010), no.~4 045120.

\bibitem{Polyakov:1975rs}
A.~M. Polyakov, ``{Compact Gauge Fields and the Infrared Catastrophe},'' {\em
  Phys. Lett. B} {\bf 59} (1975) 82--84.

\bibitem{Polyakov:1976fu}
A.~M. Polyakov, ``{Quark Confinement and Topology of Gauge Groups},'' {\em
  Nucl. Phys. B} {\bf 120} (1977) 429--458.

\bibitem{Gazit:2018vsa}
S.~Gazit, F.~F. Assaad, S.~Sachdev, A.~Vishwanath, and C.~Wang, ``{Confinement
  transition of \ensuremath{\mathbb{Z}}2 gauge theories coupled to massless
  fermions: Emergent quantum chromodynamics and SO(5) symmetry},'' {\em Proc.
  Nat. Acad. Sci.} {\bf 115} (2018), no.~30 E6987--E6995,
  \href{http://arxiv.org/abs/1804.01095}{{\tt 1804.01095}}.

\bibitem{Bi:2018xvr}
Z.~Bi and T.~Senthil, ``{Adventure in Topological Phase Transitions in 3+1 -D:
  Non-Abelian Deconfined Quantum Criticalities and a Possible Duality},'' {\em
  Phys. Rev. X} {\bf 9} (2019), no.~2 021034,
  \href{http://arxiv.org/abs/1808.07465}{{\tt 1808.07465}}.

\bibitem{Bi:2019ers}
Z.~Bi, E.~Lake, and T.~Senthil, ``{Landau ordering phase transitions beyond the
  Landau paradigm},'' {\em Phys. Rev. Res.} {\bf 2} (2020), no.~2 023031,
  \href{http://arxiv.org/abs/1910.12856}{{\tt 1910.12856}}.

\bibitem{Ma_2020}
R.~Ma and Y.-C. He, ``Emergent QCD${}_3$ quantum phase transitions of
  fractional Chern insulators,'' {\em Physical Review Research} {\bf 2} (Sept.,
  2020) \href{http://arxiv.org/abs/2003.05954}{{\tt 2003.05954}},
  \href{http://dx.doi.org/10.1103/PhysRevResearch.2.033348}{http://dx.doi.org/10.1103/PhysRevResearch.2.033348}.

\bibitem{Zou:2020hqi}
L.~Zou and D.~Chowdhury, ``{Deconfined metallic quantum criticality: A $U(2)$
  gauge-theoretic approach},'' {\em Phys. Rev. Res.} {\bf 2} (2020), no.~2
  023344, \href{http://arxiv.org/abs/2002.02972}{{\tt 2002.02972}}.

\bibitem{Wang:2021hob}
J.~Wang and Y.-Z. You, ``{Gauge enhanced quantum criticality beyond the
  standard model},'' {\em Phys. Rev. D} {\bf 106} (2022), no.~2 025013,
  \href{http://arxiv.org/abs/2106.16248}{{\tt 2106.16248}}.

\bibitem{Wang:2021vki}
J.~Wang and Y.-Z. You, ``{Gauge Enhanced Quantum Criticality Between Grand
  Unifications: Categorical Higher Symmetry Retraction},''
  \href{http://arxiv.org/abs/2111.10369}{{\tt 2111.10369}}.

\bibitem{You:2022egv}
Y.-Z. You and J.~Wang, ``{Deconfined Quantum Criticality among Grand Unified
  Theories},'' \href{http://arxiv.org/abs/2202.13498}{{\tt 2202.13498}}.

\bibitem{shor1996fault}
P.~W. Shor, ``Fault-tolerant quantum computation,'' {\em Proceedings of 37th
  conference on foundations of computer science} (1996) 56--65.

\bibitem{aharonov1997fault}
D.~Aharonov and M.~Ben-Or, ``Fault-tolerant quantum computation with constant
  error,'' {\em Proceedings of the twenty-ninth annual ACM symposium on Theory
  of computing} (1997) 176--188.

\bibitem{kitaev2003fault}
A.~Y. Kitaev, ``Fault-tolerant quantum computation by anyons,'' {\em Annals of
  Physics} {\bf 303} (2003), no.~1 2--30.

\bibitem{knill1998resilient}
E.~Knill, R.~Laflamme, and W.~H. Zurek, ``Resilient quantum computation,'' {\em
  Science} {\bf 279} (1998), no.~5349 342--345.

\bibitem{preskill1998reliable}
J.~Preskill, ``Reliable quantum computers,'' {\em Proceedings of the Royal
  Society of London. Series A: Mathematical, Physical and Engineering Sciences}
  {\bf 454} (1998), no.~1969 385--410.

\bibitem{terhal2015quantum}
B.~M. Terhal, ``Quantum error correction for quantum memories,'' {\em Reviews
  of Modern Physics} {\bf 87} (2015), no.~2 307.

\bibitem{Nielsen:1981hk}
H.~B. Nielsen and M.~Ninomiya, ``{No Go Theorem for Regularizing Chiral
  Fermions},'' {\em Phys. Lett. B} {\bf 105} (1981) 219--223.

\bibitem{Eichten:1985ft}
E.~Eichten and J.~Preskill, ``{Chiral Gauge Theories on the Lattice},'' {\em
  Nucl. Phys. B} {\bf 268} (1986) 179--208.

\bibitem{Zeng:2022grc}
M.~Zeng, Z.~Zhu, J.~Wang, and Y.-Z. You, ``{Symmetric Mass Generation in the
  1+1 Dimensional Chiral Fermion 3-4-5-0 Model},'' {\em Phys. Rev. Lett.} {\bf
  128} (2022), no.~18 185301, \href{http://arxiv.org/abs/2202.12355}{{\tt
  2202.12355}}.

\bibitem{Wang:2022ucy}
J.~Wang and Y.-Z. You, ``{Symmetric Mass Generation},'' {\em Symmetry} {\bf 14}
  (2022), no.~7 1475, \href{http://arxiv.org/abs/2204.14271}{{\tt 2204.14271}}.

\bibitem{sawyer2009calculating}
R.~Sawyer, {\em Calculating God}.
\newblock Tom Doherty Associates book. Tor Publishing Group, 2009.

\bibitem{deutsch2011beginning}
D.~Deutsch, {\em The Beginning of Infinity: Explanations That Transform the
  World}.
\newblock Penguin Publishing Group, 2011.

\end{thebibliography}\endgroup
